\documentclass[aps,prc,showpacs,showkeys,superscriptaddress,footinbib,eqsecnum]{revtex4}
\usepackage{graphicx}
\usepackage{amsmath}
\usepackage{amsfonts}
\usepackage{amssymb}
\usepackage{bm}
\newcommand{\lea}{\raisebox{-.3ex}{\small $ \
\stackrel{\textstyle <}{\sim} $ }}
\newcommand{\gea}{\raisebox{-.3ex}{\small $ \
\stackrel{\textstyle >}{\sim} $ }}

\newcommand{\beq}{\begin{equation}}
\newcommand{\eeq}{\end{equation}}
\newcommand{\beqa}{\begin{eqnarray}}
\newcommand{\eeqa}{\end{eqnarray}}
\newcommand{\be}{\begin{eqnarray}}
\newcommand{\ee}{\end{eqnarray}}
\newcommand{\eq}[1]{Eq.~(\ref{#1})}
\newcommand{\nn}{\nonumber \\ }

\date{\today}

\begin{document}

\title{Nucleon-nucleon potentials from $\Delta$-full chiral EFT and implications}

\author{Y. Nosyk}
\email{yevgenn@uidaho.edu}
\affiliation{Department of Physics, University of Idaho, Moscow, Idaho 83844, USA}
\author{D. R. Entem}
\email{entem@usal.es}
\affiliation{Grupo de F\'isica Nuclear, IUFFyM, Universidad de Salamanca, E-37008 Salamanca,
Spain}
\author{R. Machleidt}
\email{machleid@uidaho.edu}
\affiliation{Department of Physics, University of Idaho, Moscow, Idaho 83844, USA}

\begin{abstract}
We closely investigate $NN$ potentials based upon the $\Delta$-full version
of chiral effective field theory.
We find that recently constructed $NN$ potentials of this kind,
  which (when applied together with three-nucleon forces) were presented as predicting
 accurate binding energies and radii for a range of nuclei from $A=16$ to $A=132$
 and providing accurate equations of state for nuclear matter,
 yield a $\chi^2$/datum of 60 for the reproduction of the $pp$ data below 100 MeV laboratory energy.
This $\chi^2$ is more than three times what the
Hamada-Johnston potential of the year of 1962 achieved already some 60 years ago.
We perceive this historical fact as concerning
in view of the current emphasis on precision.
  We are able to trace the very large  $\chi^2$ as well as the apparent success of the potentials 
  in nuclear structure to unrealistic 
  predictions for $P$-wave states, in which the $\Delta$-full NNLO potentials are off by up to 40 times
  the NNLO truncation errors. In fact, we show that, the worse the description of the $P$-wave states, the better the predictions in nuclear structure.
 Thus, these potentials cannot be seen as the solution to the outstanding problems in
 current miscroscopic nuclear structure physics.
\end{abstract}

\maketitle

\section{Introduction}
\label{sec_intro}

One of the most fundamental aims in theoretical nuclear physics is to understand nuclear structure
and reactions in terms of the basic forces between nucleons.
As discussed in numerous review papers~\cite{ME11,EHM09,MS16,Mac17,HKK19}, the nuclear physics community presently perceives chiral effective field theory (EFT)
as the authoritative paradigm for the derivation of those forces.
This perception is based upon 
a clearly defined relationship between the fundamental theory of strong interactions, QCD,
and chiral EFT via symmetries.

Since a while, it is well established that predictive nuclear structure must include 
three-nucleon forces (3NFs), besides the usual two-nucleon force (2NF) contribution.
The advantage of chiral EFT is that it generates 2NFs and multi-nucleon forces simultaneously and 
on an equal footing. In the $\Delta$-less theory~\cite{ME11}, 3NFs occur for the first time at 
next-to-next-to-leading order (NNLO) and continue to have additional contributions in higher orders. Four-nucleon forces (4NFs) start at  next-to-next-to-next-to-leading order (N$^3$LO),
but are difficult to implement, which is why they are left out in most present-day calculations. If an explicit
$\Delta$-isobar is included in chiral EFT ($\Delta$-full theory~\cite{ORK94,ORK96,KGW98,KEM07}), then 3NF contributions start already at next-to-leading order (NLO), which leads to a smoother convergence when advancing
from leading order (LO) to NNLO. However, summing up all contributions up to NNLO leads to
very similar results for both versions of the theory~\cite{KEM07}. The convergence of both theories
beyond NNLO is expected to be very similar.

In the initial phase, 
the 3NFs were typically adjusted in
 $A=3$ and/or the $A=4$ systems and the
 {\it ab initio} calculations were driven up to the oxygen region~\cite{BNV13}.
 It turned out that for $A \lea 16$ the ground-state energies and radii are predicted about right, no matter what type of chiral or phenomenological potentials were applied (local, nonlocal, soft, hard, etc.)
 and what the details of the 3NF adjustments to few-body systems were~\cite{BNV13,Rot11,Pia18,Lon18}.
It may be suggestive to perceive the $\alpha$ substruture of $^{16}$O to be part of the explanation.

 The picture changed, when the many-body practitioners were able to move up to medium-mass nuclei (e.~g., the calcium or even the tin regions). 
 Large variations of the predictions now occurred depending on what forces were used, and cases
 of severe underbinding~\cite{Lon17} as well as of substantial overbinding~\cite{Bin14} were observed. Ever since the nuclear structure community understands that 
 the {\it ab initio} explanation of intermediate and heavy nuclei is a severe, still unsolved, problem.
 
 A seemingly successfull interaction for the intermediate mass region appears to be
 the force that is commonly denoted by ``1.8/2.0(EM)'' (sometimes
 dubbed ``the  Magic force'')~\cite{Heb11,Heb21}, which is a
 similarity renormalization group
   (SRG) evolved version of the N$^3$LO 2NF of Ref.~\cite{EM03} complemented by a NNLO 3NF adjusted to the triton binding energy and the 
 point charge radius  of $^4$He. With this force, the ground-state energies all the way up to the
 tin isotopes are reproduced perfectly---but with charge radii being on the smaller side~\cite{Sim17,Mor18}.
 Nuclear matter saturation is also reproduced reasonably well, with a slightly too high saturation density~\cite{Heb11}.
 However, these calculations are not consistently {\it ab initio}, because 
the 2NF of ``1.8/2.0(EM)'' is SRG evolved, while the 3NF is not.
 Moreover, the SRG evolved 2NF
 is used like an original force with the induced 3NFs omitted. 
 Still, this force is providing clues for how to get the intermediate and heavy mass region
 right.
 
 Thus, in the follow-up, there have been attempts to get the medium-mass nuclei under control
 by means of 
 more consistent {\it ab initio} calculations~\cite{Som20}. Of the various efforts, we will now single  out three, which demonstrate in more detail what the problems are.
 
 In Ref.~\cite{DHS19},
 recently developed soft chiral 2NFs~\cite{EMN17} at NNLO and N$^3$LO
 were picked up
 and complemented with 3NFs at NNLO and N$^3$LO, respectively, to fit the triton
 binding energy and nuclear matter saturation. These forces were then applied in
 in-medium similarity renormalization group (IM-SRG~\cite{Her16})
 calculations of finite nuclei up to $^{68}$Ni predicting underbinding and slightly too large radii~\cite{Hop19}.
 
 In a separate study~\cite{Hut20}, the same 2NFs used in Refs.~\cite{DHS19,Hop19} were employed, but with the 3NFs now adjusted
 to the triton and $^{16}$O ground-state energies.  The interactions so obtained reproduce
 accurately experimental energies and point-proton radii of nuclei up to $^{78}$Ni~\cite{Hut20}.
 However, when the 2NF plus 3NF combinations of Ref.~\cite{Hut20} are utilized in nuclear matter, then dramatic overbinding and no saturation
 at reasonable densities is obtained~\cite{SM20}.
 
 Obviously, there is a problem with achieving simultaneously reasonable results
 for nuclear matter and medium mass nuclei: In Refs.~\cite{DHS19,Hop19}, nuclear matter is saturated right, but nuclei are underbound; while in Ref.~\cite{Hut20}, nuclei are bound accurately, but nuclear matter is overbound.
 
 In recent work by the 
 G\H{o}teborg-Oak Ridge (GO) group~\cite{Eks18,Jia20},
  the authors present an NNLO model including  $\Delta$-isobars that apparently overcomes the above problem.
 With this model, the authors obtain
 ``accurate binding energies and radii for a range of nuclei from $A=16$ to $A=132$,
 and provide accurate equations of state for nuclear matter''~\cite{Jia20}.
 However, the accuracy of the $NN$ part of these interactions is not checked against $NN$ data.
  Another aspect of interest (not investigated
  in Refs.~\cite{Eks18,Jia20}) is if the inclusion of $\Delta$-degrees of freedom leads to a higher degree of softness.
 Note that the successful ``Magic'' 1.8/2.0(EM) potential is very soft
  since it is SRG evolved. 
  Moreover, a recent study~\cite{Lu18}, which investigated the essential elements of nuclear binding using nuclear lattice simulations, has come to the conclusion  that 
proper nuclear matter saturation requires a considerable amout of non-locality in the $NN$ interaction implying a high degree of softness.
 
  Thus, there is a need for a deeper understanding of the elements in the recent model by the 
   GO group~\cite{Eks18,Jia20}, and how they come together to produce the reported nuclear structure predictions.
   To gain this deeper insight, we will investigate the following issues:
 \begin{enumerate}
  \item
 What are the precision and accuracy of the $\Delta$-full $NN$ potentials developed in Ref.~\cite{Jia20}?
  In the context of chiral EFT, this amounts to asking whether the precision of the $\Delta$-full potentials is consistent with 
  the uncertainty of the chiral order at which they have been derived.
  And, is the accuracy sufficient for meaningful {\it ab initio} predictions?
  If there are problems with precison and/or accuracy, how does that impact the predictions for nuclear many-body systems?
  \item
 Does the inclusion
  of $\Delta$-isobars increase the smoothness of the interaction and, if so, how does the $\Delta$ degree of freedom accomplish that?
 \end{enumerate}
  
  This paper is organized as follows: In Sec.~II, we investigate $NN$ potentials 
  based upon $\Delta$-full chiral EFT which, in Sec.~III, are applied in nuclear matter.
  Our conclusions are summarized in Sec.~IV.

  \section{Chiral two-nucleon forces including $\Delta$-isobars}
  \label{sec_2NF}
  
  \subsection{Definition of $NN$ potentials}
  
\begin{figure}[t]\centering
\scalebox{0.55}{\includegraphics{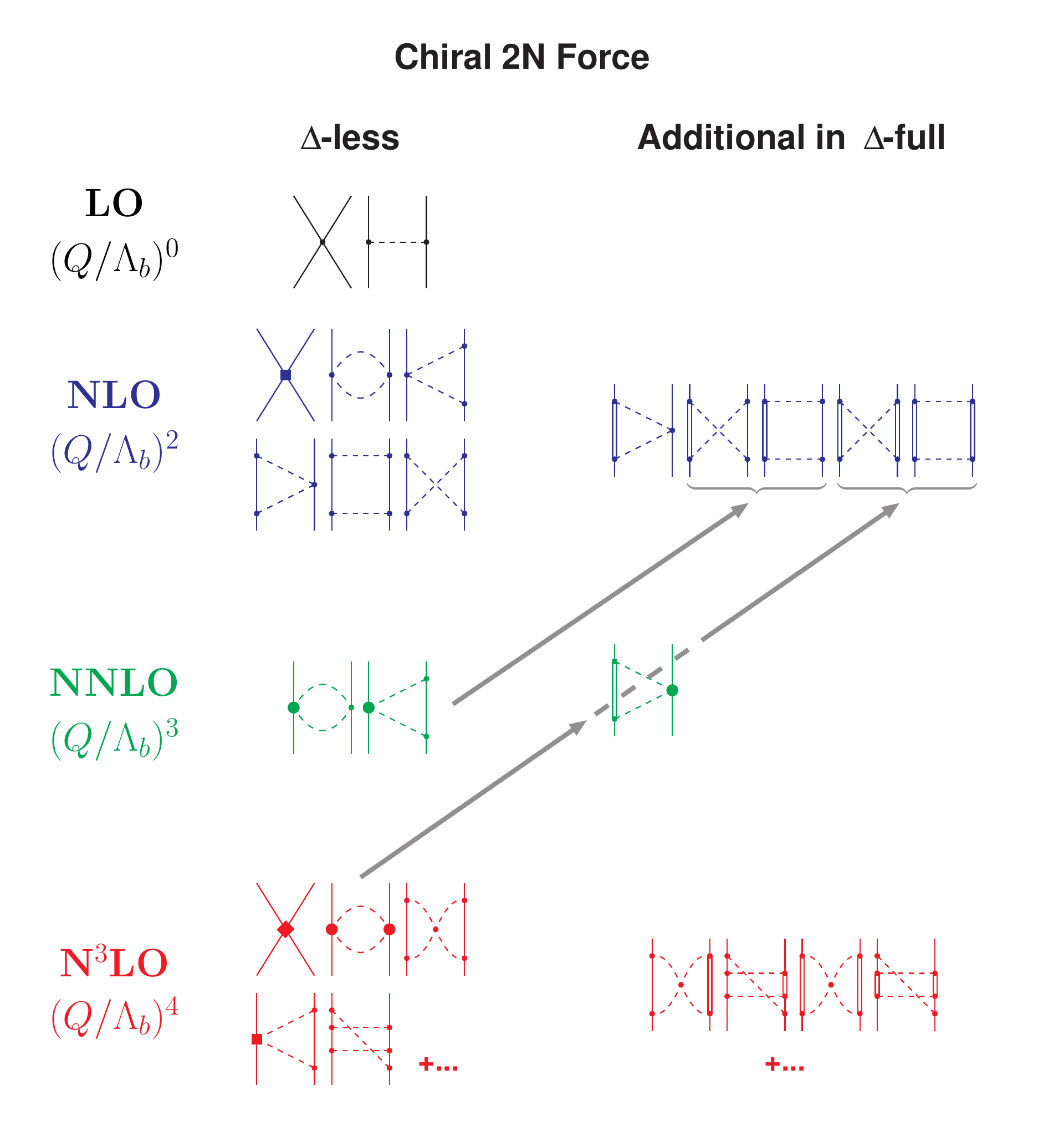}}
\vspace*{-0.5cm}
\caption{Chiral 2NF without and with $\Delta$-isobar degrees of freedom.
Arrows indicate the shift of strength when explicit $\Delta$'s are added to the theory.
Note that the $\Delta$-full theory consists of the diagrams involving $\Delta$'s
{\it plus} the $\Delta$-less ones. Solid lines
represent nucleons, double lines $\Delta$-isobars, and dashed lines pions. 
Small dots, large solid dots, solid squares, and diamonds
denote vertices of index $\delta_i= \, $ 0, 1, 2, and 4, respectively. 
$\Lambda_b$ denotes the breakdown scale.
Further explanations are
given in the text.}
\label{fig_delta_2nf}
\end{figure}

  We focus on $NN$ potentials at NNLO of the $\Delta$-full theory,
  which---following the notation introduced in Ref.~\cite{Jia20}---will be denoted by
  ``$\Delta$NNLO.''
  The diagrams to consider are displayed in Fig.~\ref{fig_delta_2nf}.
  For illustrative purposes, the figure includes also the graphs that occur at N$^3$LO.
  The powers that are associated with the various orders are calculated as follows.
  For a connected diagram of $NN$ scattering,
 the power is given by~\cite{ME11}
\begin{equation} \nu =   2L 
+ \sum_i \delta_i \, ,
\label{eq_nu} 
\end{equation}
with vertex index
\begin{equation}
\delta_i  \equiv   d_i + \frac{f_i}{2} - 2  \, ,
\label{eq_Deltai}
\end{equation}
where
$L$ denotes 
the number of loops. Moreover, for each vertex $i$,
$d_i$ is the number of derivatives or pion-mass insertions 
and $f_i$ the number of fermion fields.
The sum runs over all vertices $i$ contained in the diagram 
under consideration.

The mathematical expressions defining the potentials are given in the appendices.

We list the constants involved in the long-range parts of the potentials
(cf.\ Appendix~\ref{app_a})
  in Table~\ref{tab_basic}. These constants have the same values as used in Ref.~\cite{Jia20}.
 The $\pi N$ LECs are from the $\pi N$ analysis by Siemens {\it et al.}~\cite{Sie17}, in which
  the (redundant) subleading $\pi N \Delta$ couplings proportional to $b_3$ and $b_6$ 
  ($b_8$ in the notation of Refs.~\cite{KEM07,FM01}) are 
  removed by means of a redefinition (renormalization) of the leading order $\pi N \Delta$ 
  axial coupling $h_A$ and the subleading $\pi\pi NN$ couplings 
  $c_i$ ($i=1,2,3,4$)~\cite{Sie20}.
 
 The constants that parametrize the short-range parts of the potentials (``$NN$ contact terms,'' cf.\ Appendix~\ref{app_b}) are shown in Table~\ref{tab_ct}.

  \begin{table}[t]
\caption{Hadron masses~\cite{PDG} and pion low-energy constants~\cite{Sie17,Sie20} used throughout this work.}
\label{tab_basic}
\smallskip
\begin{tabular}{lcl}
\hline 
\hline 
\noalign{\smallskip}
  Quantity            &  \hspace{1cm} & Value \\
\hline
\noalign{\smallskip}
Charged-pion mass $m_{\pi^\pm}$ && 139.5702 MeV \\
Neutral-pion mass $m_{\pi^0}$ && 134.9766 MeV \\
Average pion-mass $\bar{m}_\pi$ && 138.0390 MeV \\
Proton mass $M_p$ && 938.2720 MeV \\
Neutron mass $M_n$ && 939.5654 MeV \\
Average nucleon-mass $\bar{M}_N$ && 938.9183 MeV \\
$\Delta$-isobar mass $M_{\Delta}$ && 1232 MeV \\
 $\Delta\equiv  M_\Delta - \bar{M}_N$ && 293.0817 MeV \\
Nucleon axial coupling constant  $g_A$ && 1.289 \\
$\pi N \Delta$ axial coupling constant  $h_A$ && 1.400 \\
Pion-decay constant $f_\pi$ && 92.2 MeV \\
$c_1$ && $-0.74$ GeV$^{-1}$\\
$c_2$ && $-0.49$ GeV$^{-1}$\\ 
$c_3$ && $-0.65$ GeV$^{-1}$\\
$c_4$ && $+0.96$ GeV$^{-1}$\\
\hline
\hline
\noalign{\smallskip}
\end{tabular}
\end{table}

\begin{table}
\caption{Partial-wave contact LECs 
for the $NN$ potentials discussed in this paper.
The  $\widetilde{C}_i$ of the zeroth order partial-wave contact terms defined in Eq.~(\ref{eq_ct0_pw}) 
are in units of $10^4$ GeV$^{-2}$ and
the $C_i$, Eq.~(\ref{eq_ct2_pw}), in $10^4$ GeV$^{-4}$.
 For SFR and regulator parameters, see Appendix~\ref{sec_nlo} and \ref{sec_reno}, respectively.
 \label{tab_ct}}
\smallskip
\begin{tabular*}{\textwidth}{@{\extracolsep{\fill}}ccccc}
\hline 
\hline 
\noalign{\smallskip}
LEC & $\Delta$NNLO(450)$_ {\bf GO}$ & $\Delta$NNLO(394)$_ {\bf GO}$
& $\Delta$NNLO(450)$_{\bf Rf}$ & $\Delta$NNLO(394)$_{\bf Rf}$
 \\
 \hline
 \noalign{\smallskip}
  $\widetilde{C}_{^1S_0}^{(pp)}$ &  --0.339111 & --0.338142 & --0.326970 & --0.327058 \\
  $\widetilde{C}_{^1S_0}^{(nn)}$ &  --0.339887 & --0.338746  & --0.3274139 & --0.32747485 \\
 $\widetilde{C}_{^1S_0}^{(np)}$ &  --0.340114 & --0.339250 & --0.32778548 & --0.32798615 \\
  $\widetilde{C}_{^3S_1}$   & --0.253950 & --0.259839 & --0.22116035 & --0.23998011  \\
   \hline
 \noalign{\smallskip}
$C_{^1S_0}$  &  2.526636 & 2.505389 & 2.238414 & 2.180000 \\
$C_{^3S_1}$   & 0.964990 & 1.002189 & 0.760000 & 0.870000 \\
 $C_{^3S_1-^3D_1}$  & 0.445743 & 0.452523 & 0.370000 & 0.435000 \\
 $C_{^1P_1}$   & --0.219498 & --0.387960 & 0.027506 & 0.027506 \\
 $C_{^3P_0}$   & 0.671908 & 0.700499 & 0.858000 & 0.892000\\ 
 $C_{^3P_1}$   &--0.915398 & --0.964856 & --0.843000 & --0.843000 \\
 $C_{^3P_2}$   &--0.895405  & --0.883122 & --0.740000 & --0.755000 \\
\hline
\hline
\noalign{\smallskip}
\end{tabular*}
\end{table}

\subsection{Predictions for two-nucleon scattering}

We will present the predictions that can be made within the $\Delta$NNLO model
in two steps. First, we will show the results obtained by the G\H{o}teborg-Oak Ridge (GO) group. 
In a second step, we will
 generate further fits of the $NN$ data within the $\Delta$NNLO model.

\subsubsection{Predictions by the GO models}

In Ref.~\cite{Jia20}, the GO group presented two $\Delta$NNLO models, which---following the GO notation---are marked by $\Delta$NNLO(450)$_ {\bf GO}$ and $\Delta$NNLO(394)$_ {\bf GO}$, where the parenthetical number denotes the value for the cutoff $\Lambda$ in units of MeV
used in the regulator function, Eq.~(\ref{eq_f}). Note that all models discussed in this paper share
the same `basic parameters' shown in Table~\ref{tab_basic}; the models differ only by the
 contact term LECs displayed in Table~\ref{tab_ct}
 (and SFR and regulator parameters, see Appendix~\ref{sec_nlo} and \ref{sec_reno}, respectively).
 The LECs listed
 in columns $\Delta$NNLO(450)$_ {\bf GO}$ and $\Delta$NNLO(394)$_ {\bf GO}$
 of Table~\ref{tab_ct}
   are from Ref.~\cite{Jia20}.

  \begin{figure}[t]
\hspace*{-0.6cm}
\scalebox{0.57}{\includegraphics{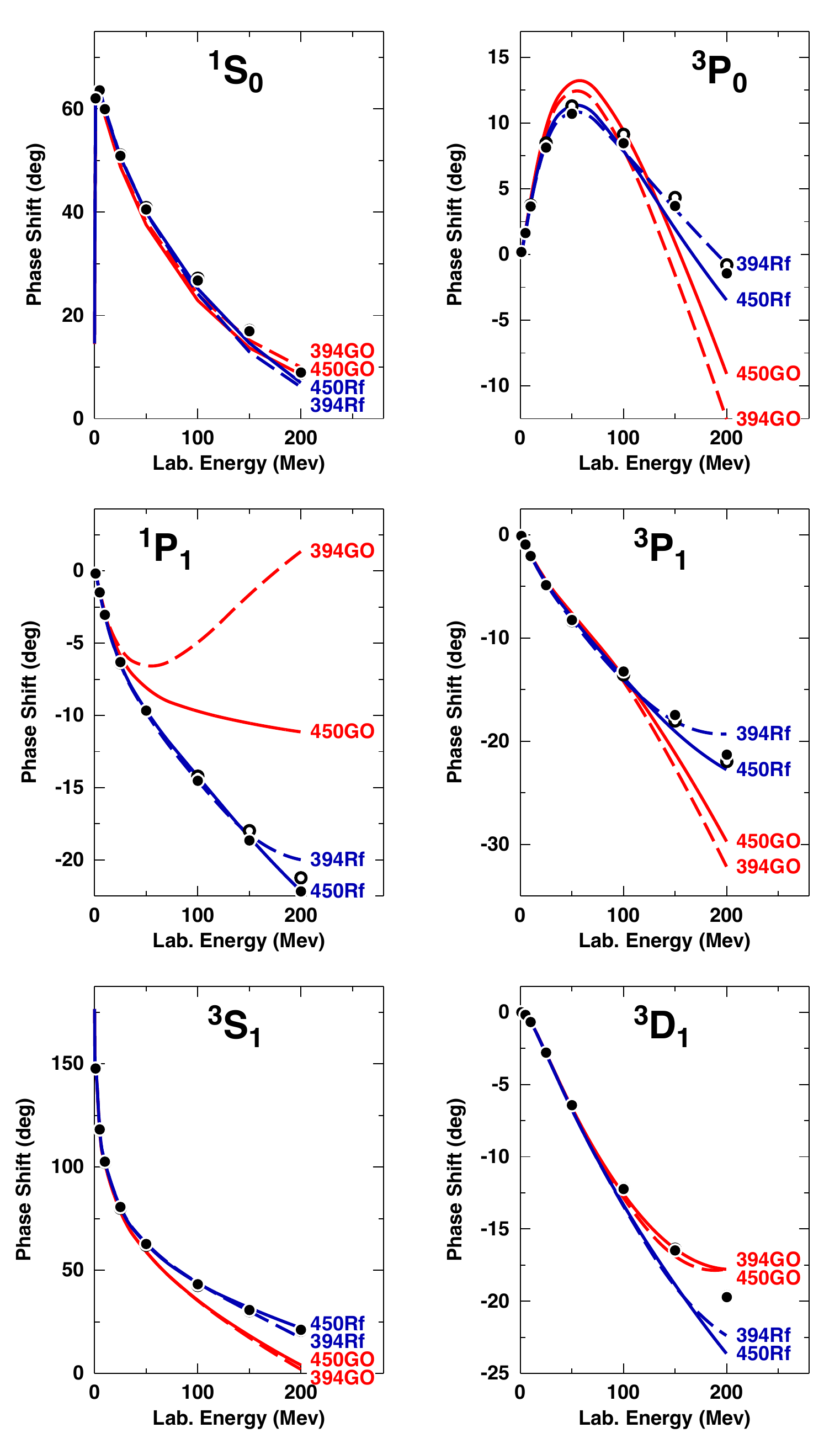}}
\hspace*{0.2cm}
\scalebox{0.57}{\includegraphics{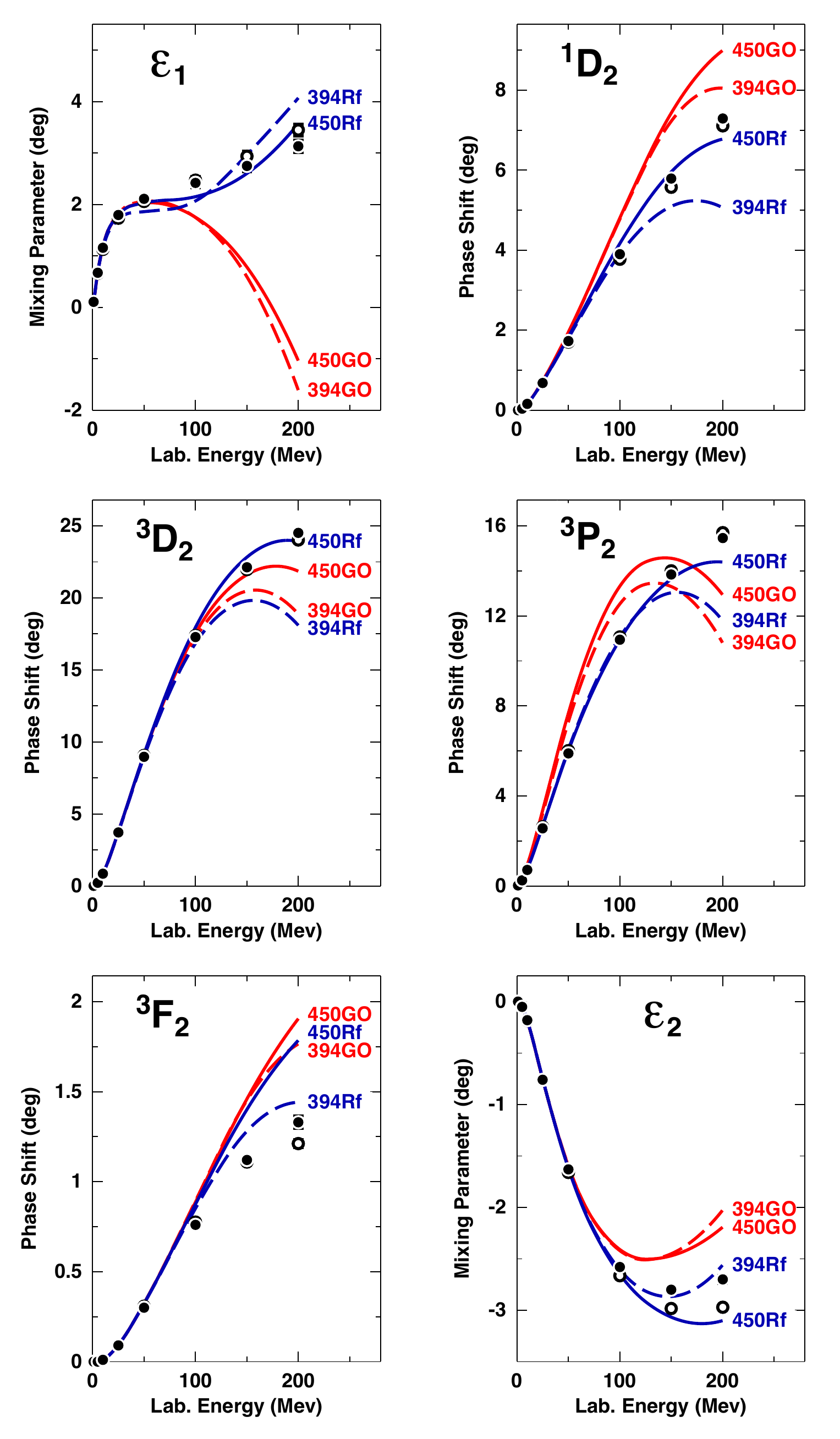}}
\vspace*{-0.3cm}
\caption{Neutron-proton phase parameters 
as predicted by the G\H{o}teborg-Oak Ridge (GO) potentials~\cite{Jia20} 
 [solid red line $\Delta$NNLO(450)$_ {\bf GO}$, dashed red 
$\Delta$NNLO(394)$_ {\bf GO}$] and by our refit (Rf) potentials  [solid blue line $\Delta$NNLO(450)$_ {\bf Rf}$, dashed blue $\Delta$NNLO(394)$_ {\bf Rf}$].
Partial waves and mixing parameters with total angular momentum $J \leq 2$ 
 are displayed for lab.\ energies up to 200 MeV.
The filled and open circles represent the results from the Nijmegen~\cite{Sto93} and the 
Granada~\cite{PAA13} $np$ phase-shift analyses, respectively.
\label{fig_ph1}}
\end{figure}

In Fig.~\ref{fig_ph1}, we display the phase parameters for neutron-proton scattering
as predicted by the GO models [solid red line $\Delta$NNLO(450)$_ {\bf GO}$, dashed red 
$\Delta$NNLO(394)$_ {\bf GO}$] and compare them with two authoritative phase-shift
analyses, namely, the Nijmegen~\cite{Sto93} and the Granada~\cite{PAA13} $np$
analyses. It is clearly seen that, above around 100 MeV
laboratory energy, the predictions deviate substantially from the
analyses in most cases.

Even though it is not uncommon to use phase shifts 
to provide a qualitative overview,
a more precise measure for the
accuracy and precision
of predictions is obtained from a direct comparison with the $NN$ data.
It is customary to state the result of such comparison
in terms of the $\chi^2$, which is obtained as outlined below.

The experimental data are broken up into
groups (sets) of data, $A$, with
$N_A$ data points and 
an experimental over-all normalization uncertainty $\Delta n_A^{exp}$. 
For datum $i$,
$x^{exp}_i$ is the experimental value,
$\Delta x^{exp}_i$ the experimental uncertainty, and
$x^{mod}_i$ the model prediction.
When fitting the data of group $A$ by a model (or a phase shift
solution), the over-all normalization, $n_A^{mod}$, is floated and finally
chosen such as to minimize the $\chi^2$ for this group.
The $\chi^2$ is then calculated from~\cite{Ber88}
\begin{equation}
\chi^2= \sum_A \left\{ \sum^{N_A}_{i=1} \left[
\frac{n_A^{mod} x^{mod}_i-x^{exp}_i}{\Delta x^{exp}_i}
\right]^2
+ \left[ \frac{n_A^{mod} -1}{\Delta n_A^{exp}} \right]^2
\right\} \; ;
\label{eq_chi2}
\end{equation}
that is, the over-all normalization of a group is treated as
an additional datum.
For groups of data without normalization uncertainty
($\Delta n_A^{exp}=0$), $n_A^{mod}=1$ is used and the second
term on the r.h.s.\ of Eq.~(\ref{eq_chi2}) is dropped.
The total number of data is
\begin{equation}
N_{dat}=N_{obs}+N_{ne}
\end{equation}
where $N_{obs}$ denotes the total number of measured data points (observables), i.~e.,
$N_{obs}=\sum_A N_A$; and
$N_{ne}$ is the number of experimental normalization
uncertainties.
We state results in terms of $\chi^2/N_{dat} \equiv  \chi^2/$datum, where we use,
in general, for the experimental data the
 2016 $NN$ base which is defined in Ref.~\cite{EMN17}.

  \begin{table}
\caption{$\chi^2/$datum for the reproduction of the $NN$ data by the  G\H{o}teborg-Oak Ridge (GO) potentials and by our refit (Rf) potentials. The Hamada-Johnston potential~\cite{HJ62} is included for comparison.}
\label{tab_chi}
\smallskip
\begin{tabular*}{\textwidth}{@{\extracolsep{\fill}}cccccc}
\hline 
\hline 
\noalign{\smallskip}
      & Hamada-Johnston \\
 Bin (MeV) & Potential & $\Delta$NNLO(450)$_ {\bf GO}$ & $\Delta$NNLO(394)$_ {\bf GO}$
 & $\Delta$NNLO(450)$_{\bf Rf}$ & $\Delta$NNLO(394)$_{\bf Rf}$  \\
                 & of 1962~\cite{HJ62,foot1,SS93} \\
\hline
\noalign{\smallskip}
\multicolumn{6}{c}{\bf proton-proton} \\
0--100 & 19.6   & 60.7 & 34.3 & 2.07 &1.87 \\
0--200 & 13.8 & 46.3 & 39.7 & 5.39 & 10.7 \\
\hline
\noalign{\smallskip}
\multicolumn{6}{c}{\bf neutron-proton} \\
0--100 &  &  5.87 & 8.58 & 1.27 & 1.20  \\
0--200 &  & 14.2 & 26.2  & 2.23 & 9.60 \\
\hline
\noalign{\smallskip}
\multicolumn{6}{c}{\boldmath $pp$ plus $np$} \\
0--100 &  & 28.8 & 19.3 & 1.59  & 1.47 \\
0--200 &  & 29.6 & 32.6 & 3.71 & 10.1 \\
\hline
\hline
\noalign{\smallskip}
\end{tabular*}
\end{table}

In Table~\ref{tab_chi}, we show the $\chi^2/$datum for the two  G\H{o}teborg-Oak Ridge potentials,
$\Delta$NNLO(450)$_ {\bf GO}$ and $\Delta$NNLO(394)$_ {\bf GO}$, for $pp$ scattering,
$np$ scattering, and a combination of both  for the lab.\ energy intervals 0--100 and 0--200 MeV. 
In the case of the $\Delta$NNLO(450)$_ {\bf GO}$ potential,
the over-all $\chi^2/$datum for $pp$ plus $np$ is about 30 for all intervals considered, while for 
$\Delta$NNLO(394)$_ {\bf GO}$ the
$\chi^2/$datum lies around 20 to 30. 
The $pp$ $\chi^2$ for the
interval 0--100 MeV
is particularly concerning, because one should expect lower $\chi^2$ for lower
energies, whereas, in the case of $\Delta$NNLO(450)$_ {\bf GO}$, the interval of lowest energy has the highest $\chi^2$.
 The reason for this anomaly can be traced to 
the $P$-wave phase-shifts
around 50 MeV (cf.\ Table~\ref{tab_ph} and Fig.~\ref{fig_ph4}, below) which---as we will demonstrate in Sec.~\ref{sec_NM}---have a dramatic impact on nuclear matter predictions. 
Notice that problems at low energies
cannot be well identified from global phase-shift plots (cf.\ Fig.~\ref{fig_ph1}), which
corroborates the limited value of phase-shift figures and underscores the importance 
of the $\chi^2$ for the fit of the experimental data.

To put the afore-mentioned $\chi^2$ values into perspective, we include in Table~\ref{tab_chi}
the $pp$ $\chi^2$ of the first semi-quantitative $NN$ potential constructed
in the history of nuclear forces: the
Hamada-Johnston potential of 1962~\cite{HJ62}.
This old-timer yields a $pp$ $\chi^2$/datum
of 13.8 for the interval 0--183 MeV~\cite{HJ62,foot1,SS93}.
Thus, the $\chi^2$/datum of 46.3 produced by the $\Delta$NNLO(450)$_ {\bf GO}$ potential is 
more than three times
larger than the one of the 60-year-old potential.
In fact, none of the historical $NN$ potentials listed in
Table II of Ref.~\cite{SS93}
has a $\chi^2$ as large as the one of the GO potentials of 2020.
Clearly, this is problematic, especially considering that high precision 
is becoming an increasingly important feature for
current advances and goals in {\it ab initio}
nuclear structure physics~\cite{INT21}.
  
  In addition to the above historical perspective, it is important to convey some clear physics arguments.
  Contemporary $NN$ potentials developed within the well-defined framework of an EFT must satisfy specific criteria. The EFT is organized order by order with an appropriate expansion parameter and, consequently,
  the precision of the predictions can be estimated---being dictated by the truncation error at the order under consideration.
  
  The expansion parameter $Q$ is given by~\cite{EKM15}
  \begin{equation}
  Q = \max \left\{ \frac{m_\pi}{\Lambda_b}, \; \frac{p}{\Lambda_b} \right\} \,,
  \end{equation}
  where $p$ is the characteristic center-of-mass (cms) momentum scale and $\Lambda_b$
  the so-called breakdown scale for which we choose a value of 700 MeV,
  consistent with the investigations of Ref.~\cite{Fur15}. The truncation error at NNLO is then 
  determined to be~\cite{EKM15}
   \begin{equation}
  \Delta X_{\rm NNLO}(p) = \max \left\{ Q^4 \times \left| X_{\rm LO}(p) \right|, \;\;
  Q^2 \times \left| X_{\rm LO}(p) - X_{\rm NLO}(p) \right|, \;\;
  Q \times \left| X_{\rm NLO}(p) - X_{\rm NNLO}(p) \right|
   \right\} \,,
   \label{eq_error}
  \end{equation}
  where $X_{\rm NNLO}(p)$ denotes the NNLO prediction for observable $X(p)$, etc..
  Since, in the $\Delta$-full theory, the difference between NLO and NNLO is very small, 
  the third term in the curly bracket is most likely not the maximum.
  Concerning the remaining two terms, let us start with the first term,
  $Q^4 \times \left| X_{\rm LO}(p) \right|$. 
  Assuming that $\left| X_{\rm LO}(p) \right|$ is of the size of
  the observable under consideration, then $Q^4$ represents the (relative) truncation error suggested by the first term.
  Since $Q$ is momentum dependent, let us consider two energy ranges:
  A low energy range ($\approx 0-100$ MeV) where $Q=m_\pi/\Lambda_b=0.2$ and an intermediate energy range ($\approx 100-200$ MeV) around a
  lab.\ energy of 150 MeV ($p=265$ MeV/c) implying $Q=p/\Lambda_b=0.4$.
  For these two energy ranges, we have $Q^4 \approx$ 
  0.002 and 0.03; or 0.2\% and 3\%, respectively.
  When calculating error estimates for the phase shifts shown in Table~\ref{tab_ph}, below, we made
  the experience that the second term in the curly bracket of Eq.~(\ref{eq_error}), namely
  the $Q^2 \times \left| X_{\rm LO}(p) - X_{\rm NLO}(p) \right|$ term , is in general the largest
 one and as a rule of thumb about twice the $Q^4$ term. Therefore, to be on the conservative side,
  we double the naive estimates and assume truncation errors of
  0.4\% and 6\% for the lab.\ energy intervals $0-100$ and $100-200$ MeV, respectively. 
  
  To make connection with the $\chi^2$ formula, Eq.~(\ref{eq_chi2}),
  one may identify
  $\Delta X_{\rm NNLO}(p) \approx \left|
  (n_A^{mod} x^{mod}_i-x^{exp}_i)/x^{exp}_i \right|$
  for pieces of data $x^{exp}_i$ in the energy range characterized by the cms momentum $p$.
  Thus, to estimate the $\chi^2$, one needs an idea of how the truncation error compares to typical experimental errors.
  
  Going over the comprehensive $pp$ data base of Ref.~\cite{Ber90} reveals that,
  for low energies, experimental errors around 0.2-0.4\% are not uncommon.
  At intermediate energies,
  the experimental errors move up to typically 2-4\%
   for the $pp$ as well as the $np$ data~\cite{Ber90,Sto93}.
  Thus, $\chi^2$/datum around 1-2 for low energies and around 2-5 for the higher energy intervall
  are consistent with the estimated truncation error at NNLO. 
  More compelling evidence is provided by actual calculations. For the $\Delta$-less theory,
  systematic order-by-order calculations with minimized $\chi^2$ have been conducted in 
  Refs.~\cite{EMN17,RKE18}. In the case of the NNLO potential of Ref.~\cite{EMN17},
  $\chi^2$/datum of 1.7 and 3.3 are generated for the intervals
  0-100 and 0-190 MeV for the combined $np$ plus $pp$ data.
  These results are in line with our above estimates based upon the truncation error
  at NNLO, and indicate that a $\chi^2$/datum $\approx 30$ is inconsistent 
  with the precision at NNLO.
  
   Low energy scattering parameters and deuteron properties are shown in Table~\ref{tab_lep}
  and \ref{tab_deu}, respectively, which reveal further inaccuracies in the GO potentials.
  
  Some important phase shifts and their NNLO truncation uncertainties are displayed in Table~\ref{tab_ph}, 
from which one must conclude that the phase shift predictions by the GO potentials 
are off by 40 times the truncation error in some cases.

 \begin{table}
\caption{Scattering lengths ($a$) and effective ranges ($r$) in units of fm as predicted by 
the G\H{o}teborg-Oak Ridge (GO) potentials and by our refit (Rf) potentials.
($a_{pp}^C$ and $r_{pp}^C$ refer to the $pp$ parameters in the presence of
the Coulomb force. $a^N$ and $r^N$ denote parameters determined from the
nuclear force only and with all electromagnetic effects omitted.)
\label{tab_lep}}
\smallskip
\begin{tabular*}{\textwidth}{@{\extracolsep{\fill}}cccccc}
\hline 
\hline 
\noalign{\smallskip}
       & $\Delta$NNLO(450)$_ {\bf GO}$ & $\Delta$NNLO(394)$_ {\bf GO}$ 
 & $\Delta$NNLO(450)$_{\bf Rf}$ & $\Delta$NNLO(394)$_{\bf Rf}$ & Empirical       \\
\hline
\noalign{\smallskip}
\multicolumn{6}{c}{\boldmath $^1S_0$} \\
$a_{pp}^C$  &--7.8929 & --7.8190 & --7.8153 & --7.8150 & --7.8196(26)~\cite{Ber88} \\
  &&&&& --7.8149(29)~\cite{SES83} \\
$r_{pp}^C$  &  2.870   & 2.865 & 2.761 & 2.732 &  2.790(14)~\cite{Ber88}  \\
  &&&&& 2.769(14)~\cite{SES83} \\
$a_{pp}^N$  & --17.670 & --17.377 & --17.824 & --17.901 & \\
$r_{pp}^N$  & 2.953 & 2.944 & 2.821 & 2.791 & \\
$a_{nn}^N$  & --19.382 & --18.723 & --18.950 & --18.950 & --18.95(40)~\cite{Gon06,Che08} \\
$r_{nn}^N$  & 2.919 & 2.916 & 2.800 & 2.772 &  2.75(11)~\cite{MNS90} \\
$a_{np}  $  &--23.560 & --23.504 & --23.738 & --23.738 & --23.740(20)~\cite{Mac01} \\
$r_{np}  $  &  2.813 & 2.797 & 2.686 & 2.661 & [2.77(5)]~\cite{Mac01}   \\
\hline
\noalign{\smallskip}
\multicolumn{6}{c}{\boldmath $^3S_1$} \\
$a_t$     &  5.458 & 5.463 & 5.422 & 5.418 & 5.419(7)~\cite{Mac01}  \\
$r_t$     &  1.820 & 1.820 & 1.757 & 1.751 & 1.753(8)~\cite{Mac01}  \\
\hline
\hline
\noalign{\smallskip}
\end{tabular*}
\end{table}

\begin{table}
\small
\caption{Deuteron properties as predicted by
  the $NN$ potentials of this study.
(Binding energy $B_d$, asymptotic $S$ state $A_S$,
asymptotic $D/S$ state $\eta$, structure radius $r_{\rm str}$,
quadrupole moment $Q$, $D$-state probability $P_D$; the predicted
$r_{\rm str}$ and $Q$ are without meson-exchange current contributions
and relativistic corrections.)
\label{tab_deu}}
\smallskip
\begin{tabular*}{\textwidth}{@{\extracolsep{\fill}}llllll}
\hline 
\hline 
\noalign{\smallskip}
   &  $\Delta$NNLO(450)$_ {\bf GO}$ & $\Delta$NNLO(394)$_ {\bf GO}$  
   & $\Delta$NNLO(450)$_{\bf Rf}$ & $\Delta$NNLO(394)$_{\bf Rf}$ 
   & Empirical$^a$ \\
\hline
\noalign{\smallskip}
$B_d$ (MeV) &
 2.233403 & 2.227450 & 2.224575 & 2.224575
  & 2.224575(9) \\
$A_S$ (fm$^{-1/2}$) &
  0.8954 & 0.8943 & 0.8856 & 0.8849
 & 0.8846(9)  \\
$\eta$         & 
 0.0253 & 0.0254 & 0.0257 & 0.0256
& 0.0256(4) \\
$r_{\rm str}$ (fm)   & 
 1.986 & 1.988 & 1.969 & 1.969
 & 1.97507(78) \\
$Q$ (fm$^2$) &
 0.268 & 0.267 & 0.272 & 0.267
& 0.2859(3)  \\
$P_D$ (\%)    & 
 3.12 & 2.97 & 4.16 & 3.49
 & --- \\
\hline
\hline
\noalign{\smallskip}
\end{tabular*}
\footnotesize
$^a$See Table XVIII of Ref.~\cite{Mac01} for references;
the empirical value for $r_{\rm str}$ is from Ref.~\cite{Jen11}.\\
\end{table}

  \subsubsection{Accurate fits for $\Delta$NNLO models}
  \label{sec_refit}
  
  In the next step, we have constructed 
   $\Delta$NNLO models with improved fits---for the purpose of explicitly checking out
  whether, within the $\Delta$-full theory, we can achieve
  $\chi^2$ that are consistent with the above estimates
  and the $\chi^2$ obtained in Ref.~\cite{EMN17} for the $\Delta$-less theory.
We have dubbed our refits $\Delta$NNLO(450)$_ {\bf Rf}$ 
and $\Delta$NNLO(394)$_ {\bf Rf}$
(where ``Rf'' stands
for Refit).  The parameters of the refits are listed in Table~\ref{tab_ct}~\cite{foot3}
and
the $\chi^2$/datum are shown in Table~\ref{tab_chi}. 
The phase shifts are displayed in Fig.~\ref{fig_ph1} by the blue solid and blue dashed lines.
The conclusion is that,
within the $\Delta$-full theory, fits can be achieved that are of the same
quality as in the $\Delta$-less theory and
consistent with 
the truncation error (cf.\ also Table~\ref{tab_ph}).
In the case of the very soft cutoff of 394 MeV, cutoff artefacts are obviously showing up already below 200 MeV, which is not unexpected.

\section{Nuclear matter}
\label{sec_NM}

\begin{figure}[t]\centering
\scalebox{0.9}{\includegraphics{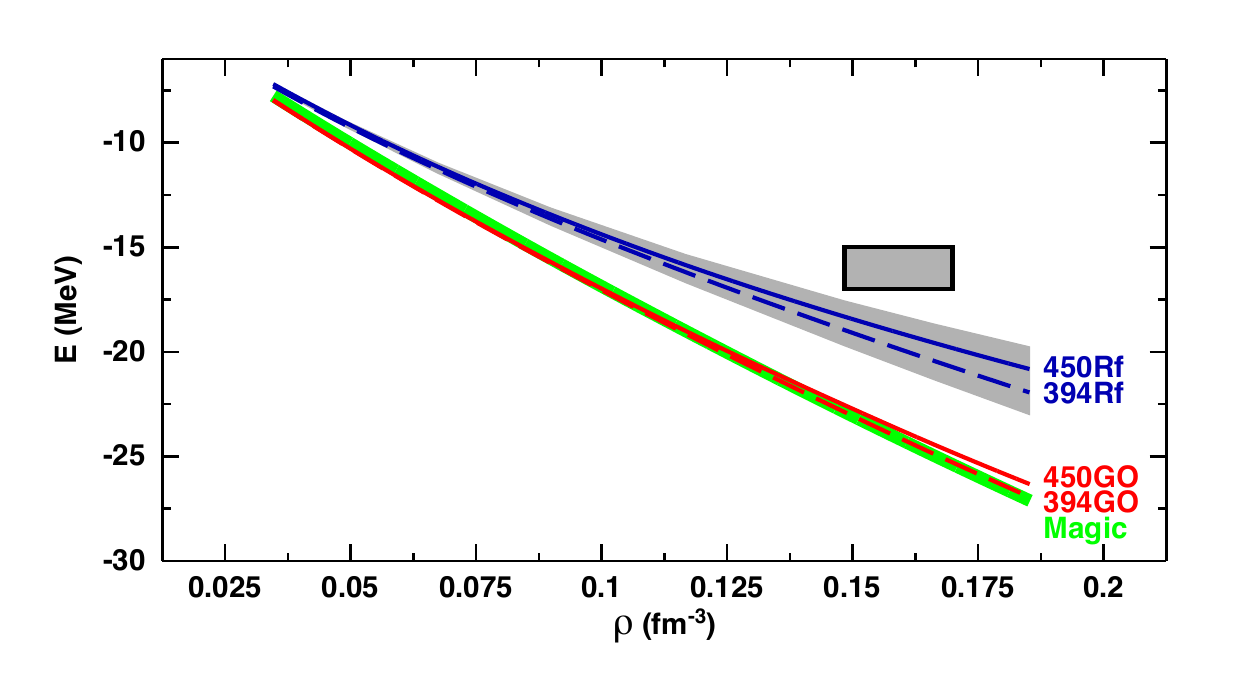}}
\vspace*{-0.5cm}
\caption{Energy per nucleon in symmetric nuclear matter, $E$, as a function of density, $\rho$, 
as generated by some two-body forces. Notation for the $\Delta$NNLO 
potentials as in Fig.~\ref{fig_ph1}.
 Magic (solid green line) refers to the 1.8/2.0(EM) potential of Ref.~\cite{Heb11}.
 The shaded band includes the theoretical uncertainties associated with the
 predictions by the Rf potentials (blue lines)~\cite{foot2}.
 Note that this shaded band also covers the predictions by the $\Delta$-less NNLO and N$^3$LO 
 potentials of Ref.~\cite{EMN17} applied in Ref.~\cite{Hop19} to intermediate-mass nuclei.
 The grey box outlines the area where nuclear saturation is expected to occur.}
\label{fig_NM}
\end{figure}

 The attempts to explain nuclear matter saturation have a long history~\cite{Bet71,Mac89}.
 The modern view is that the 3NF is essential to obtain saturation~\cite{Bog05,Heb21}.
 In this scenario, the 2NF substantially overbinds nuclear matter, while the 3NF 
 contribution is
 repulsive and strongly density-dependent leading to saturation at the appropriate
 energy and density~\cite{Mac19}. 
 Recent example can be found in the work of Ref.~\cite{DHS19}, where chiral 2NFs at NNLO and N$^3$LO
 are complemented with chiral 3NFs of the corresponding orders to saturate nuclear
 matter around its empirical values.
 
 Besides nuclear matter, there is also the problem of the binding energies of intermediate-mass
 nuclei. When the 2NF+3NF combinations of Ref.~\cite{DHS19} were applied 
 in IM-SRG calculations of finite nuclei up to the nickel isotops, underbinding 
 of the ground state energies was obtained~\cite{Hop19}.
 
 On the other hand, also in Ref.~\cite{Heb11}, 2NF+3NF combinations were developed;
 in particular, the force known as 1.8/2.0(EM) or Magic, which saturates
 nuclear matter properly and reproduces the groundstate energies
 of nuclei up to the tin region correctly~\cite{Sim17,Mor18}.
 
 What is the difference between the two cases?

   \begin{figure}[t]
\scalebox{0.7}{\includegraphics{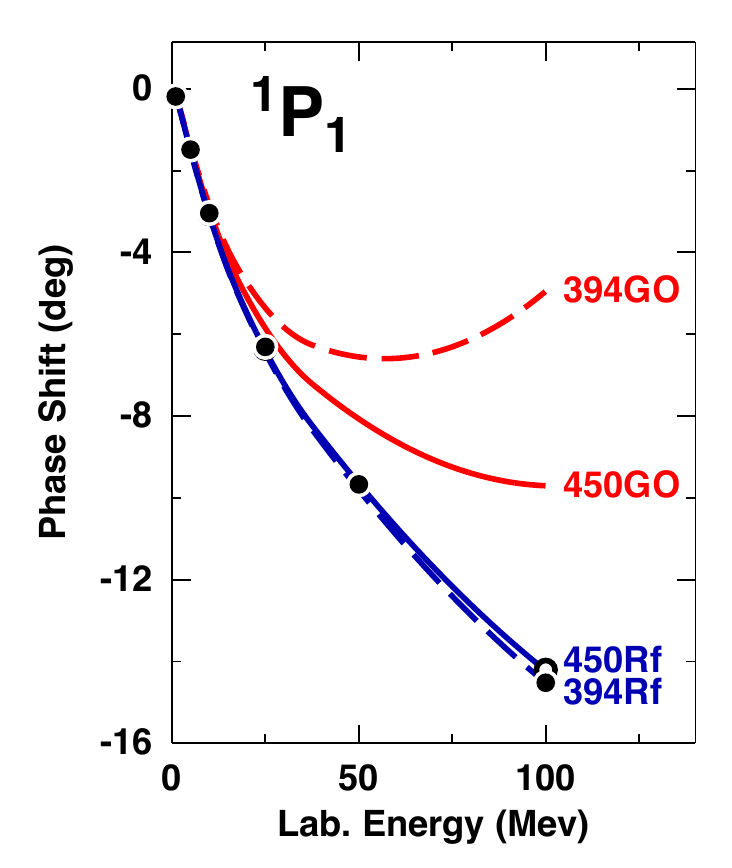}}
\scalebox{0.7}{\includegraphics{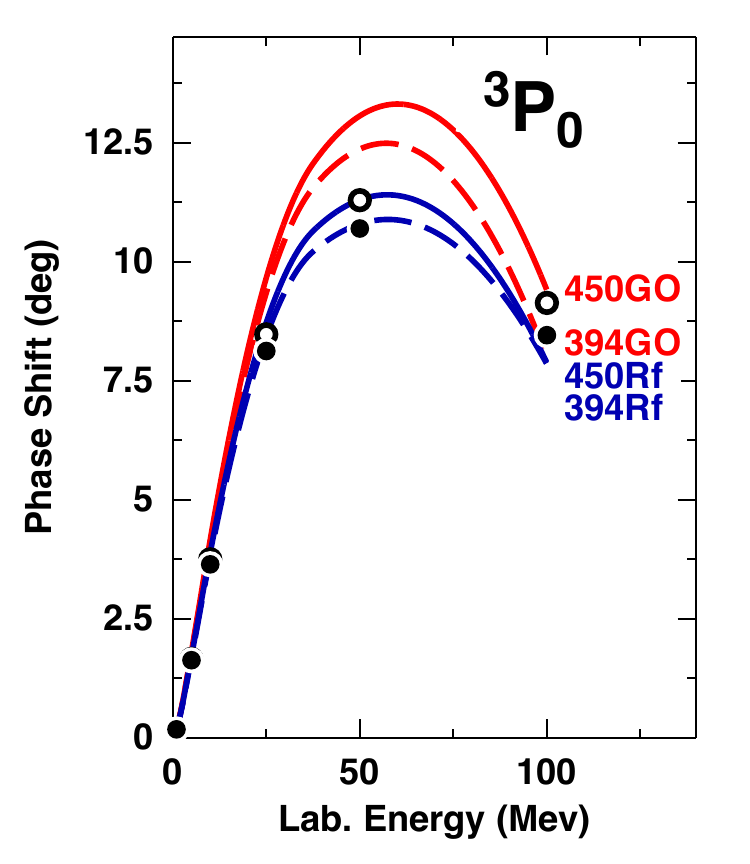}}
\scalebox{0.7}{\includegraphics{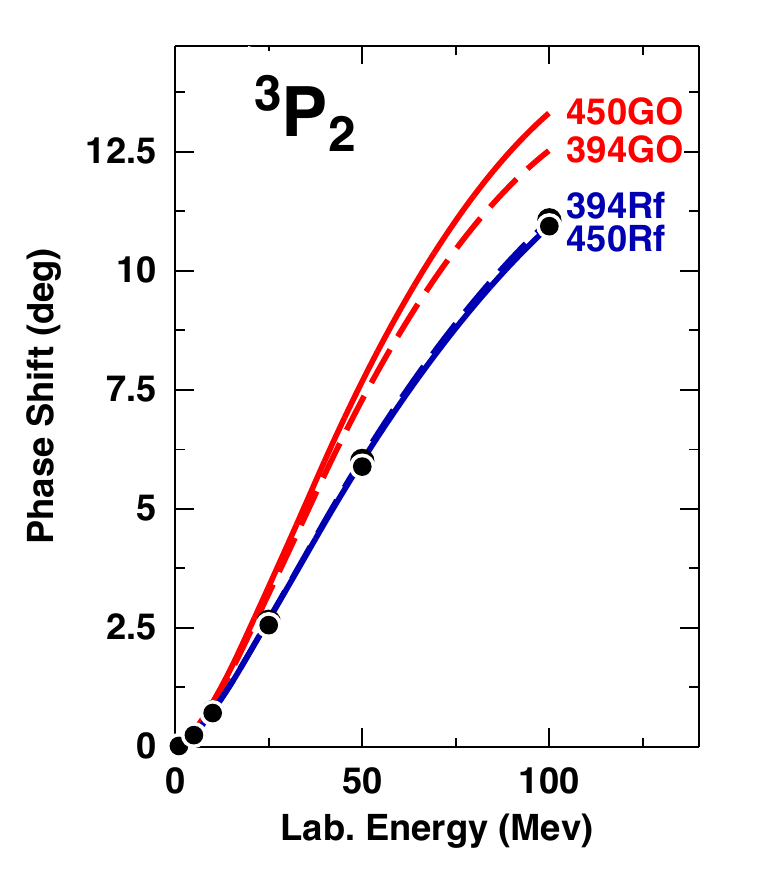}}
\caption{Neutron-proton phase shifts below 100 MeV for three critical $P$-waves. Notation as in
Fig.~\ref{fig_ph1}.}
\label{fig_ph4}
\end{figure}

 As it turns out
 the crucial difference between the two cases is to be found in the 2NF part of the forces. 
 We demonstrate this in Fig.~\ref{fig_NM}, where we show the 2NF contribution
 to nuclear matter from the 1.8/2.0(EM) force (denoted by Magic, solid green line).
 On the other hand, the  2NF contribution
 to nuclear matter  from
 the 2NFs applied in Ref.~\cite{Hop19} are located within the shaded band in Fig.~\ref{fig_NM}.
 Even though in both cases nuclear matter is  overbound,  Magic overbinds considerable more than the
 NNLO and N$^3$LO forces from Ref.~\cite{EMN17}  applied in Ref.~\cite{Hop19} to intermediate-mass nuclei.
 This shows that a considerable overbinding of nuclear matter by the 2NF
 is necessary to correctly bind intermediate-mass nuclei~\cite{SM20}, 
 when 3NFs at NNLO are applied.
 
  Next, we turn to the nuclear matter properties as predicted by the $\Delta$-full  2NFs
 discussed in this paper. 
 We apply the particle-particle ladder approximation [Brueckner-Hartree-Fock (BHF)]
 for nuclear matter.
 We have compared our nuclear matter results for the Magic 2NF with the results obtained in Ref.~\cite{Heb11} where many-body perturbation theory (MBPT) is used and obtain the same
 within $\pm 0.5$ MeV for all densities displayed in Ref.~\cite{Heb11}. Moreover, we have also compared our BHF results with the MBPT calculations of Ref.~\cite{DHS19} (for the $\Lambda = 450$ MeV potentials) achieving a similar agreement. From this we conclude that our BHF method  for nuclear matter is as reliable as the presently more popular MBPT method for soft potentials.

 The predictions by the original GO potentials,
 $\Delta$NNLO(450)$_ {\bf GO}$ and $\Delta$NNLO(394)$_ {\bf GO}$, are shown in 
 Fig.~\ref{fig_NM} by the red solid and dashed curves, respectively.
 These predictions are right on the  Magic curve, which explains the results of
 the GO potentials for nuclei up to $A=132$~\cite{Jia20}, 
 similar to  what happens with Magic~\cite{Sim17,Mor18}.
 
 On the other hand, in the previous section we have identified serious problems with the accuracy of the GO $NN$ potentials. For that reason, 
  in Sec.~\ref{sec_refit} we refitted these potentials, generating the Rf versions
 $\Delta$NNLO(450)$_{\bf Rf}$ and $\Delta$NNLO(394)$_{\bf Rf}$, which are as accurate as
 expected at NNLO.
 The nuclear matter predictions by the Rf potentials are shown by the blue solid
 and dashed curves in  Fig.~\ref{fig_NM}, together with their theoretical uncertainties represented by the shaded band.
 It is seen that
 the refitted potentials are less attractive than the original GO versions.
 In fact, their nuclear matter properties are
 very similar to the ones of the 2NFs used in Ref.~\cite{Hop19} and, thus,
 they will most likely not produce the same results as the original GO potentials and,
 rather, produce underbinding in intermediate-mass nuclei.
 
 The question we wish to address is then why, after refit to proper accuracy, the $\Delta$NNLO
 potentials lost attraction. As discussed, the main issues with the GO potentials
 are found in the $P$-waves at low energy. We demonstrate this 
 in Fig~\ref{fig_ph4}, where we show, for the three most important $P$-waves, the phase shifts below 100 MeV as
 predicted by the original GO potentials (red lines) and and the refit versions in comparison with authoritative phase
 shift analyses.

 \begin{table}
 \footnotesize
\caption{$np$ phase-parameters, $\delta$ (in degrees), for selected states and laboratory energies, $T_{\rm lab}$,
as predicted by 
the G\H{o}teborg-Oak Ridge (GO) potentials and by our refit (Rf) potentials,
for which nuclear matter predictions are
shown in Table~\ref{tab_NM}.  
Empirical values are the averages of the  Nijmegen~\cite{Sto93} and the 
Granada~\cite{PAA13} analyses. 
$\Delta \delta$ is the magnitude of the difference between a prediction and the empirical value.
$\Delta \delta_{\rm NNLO}$ denotes the theoretical uncertainty calculated according
to Eq.~(\ref{eq_error})~\cite{foot2}.
}
\label{tab_ph}
  \smallskip
\begin{tabular*}{\textwidth}{@{\extracolsep{\fill}}cccccccccccc}
\hline
\hline
\noalign{\smallskip}
 State & $\frac{T_{\rm lab}}{\rm MeV}$ & Empirical 
       & \multicolumn{2}{c}{$\Delta$NNLO(450)$_ {\bf GO}$} 
       & \multicolumn{2}{c}{$\Delta$NNLO(394)$_ {\bf GO}$} 
       & \multicolumn{2}{c}{$\Delta$NNLO(450)$_{\bf Rf}$} 
       & \multicolumn{2}{c}{$\Delta$NNLO(394)$_{\bf Rf}$}  
  &  $\Delta \delta_{\rm NNLO}$ \\
       \cline{4-5}  \cline{6-7}   \cline{8-9}   \cline{10-11}
       &       &       
    & $\delta$  &  $\frac{\Delta \delta }{ \Delta \delta_{\rm NNLO}}$ 
        & $\delta$  &  $\frac{\Delta \delta }{ \Delta \delta_{\rm NNLO}}$ 
            & $\delta$  &  $\frac{\Delta \delta }{ \Delta \delta_{\rm NNLO}}$ 
                & $\delta$  &  $\frac{\Delta \delta }{ \Delta \delta_{\rm NNLO}}$ 
\\
\hline
\hline
\noalign{\smallskip}
$^1P_1$ &  50  & --9.67(5)  &  --8.07  &  22.9  &  --6.56  &  44.4  &  --9.65  &  0.3  & --9.81 & 2.0 & 0.07     \\
$^3P_0$ & 50 & 11.00(5) & 13.07& 2.8 & 12.37 & 1.8 & 11.30 & 0.4 & 10.80 & 0.3 & 0.75 \\
$^3P_2$ & 50 & 5.95(1) & 7.68 & 7.5 & 7.29 & 5.8 & 6.01 & 0.3 & 6.07 & 0.5 & 0.23 \\
$^3S_1$ & 50 & 62.47 (6) & 59.00 & 9.1 & 59.00 & 9.1 & 62.79 & 0.9 & 63.19 & 1.9 & 0.38 \\
$\epsilon_1$ & 150 & 2.84(7) & 0.78 & 5.3 & 0.59 & 5.8 & 2.60 & 0.6 & 2.96 & 0.3 & 0.39 \\
\hline
\hline
\end{tabular*}
\end{table}

 To further quantify the discrepancies, we provide in Table~\ref{tab_ph}
 numerical values for $np$ phase-shifts at 50 MeV lab.\ energy for the three $P$-waves
 of interest as predicted by 
 the original GO potentials, the refit potentials, and the phase shift analyses (`Empirical').
 We also provide the NNLO truncation error for the phase shifts, 
 $\Delta \delta_{\rm NNLO}$~\cite{foot2}, and state the discrepancies in the fits, $\Delta \delta$,
 in terms of multiples of the truncation errors, ${\Delta \delta }/{ \Delta \delta_{\rm NNLO}}$.
 For the GO potentials, the discrepancies are on average about ten truncation errors,
 and can be as large as 44 truncation errors.
 For the refit potentials, the discrepancies are typically around one truncation error or less -- as expected
  for a properly converging EFT, whose predictions at each order should
 agree with experiment
within the theoretical uncertainty (truncation error) at the given order.

 \begin{table}
\caption{Energy contributions per nucleon to
symmetric nuclear matter from two-body forces at a density equivalent to
  a Fermi momentum $k_F=1.35$ fm$^{-1}$ 
  as obtained in the non-perturbative particle-particle ladder approximation.
  The G\H{o}teborg-Oak Ridge (GO) $NN$ potentials and our refit (Rf) potentials
  are applied. Moreover,
  Magic refers to the 1.8/2.0(EM) $NN$ potential of Ref.~\cite{Heb11}.
  $U$ represents the total potential energy per nucleon, $T$ the kinetic energy, and $E$ the total energy per nucleon,
  $E=T+U$.
  $U(...)$ denotes the potential energy contribution per nucleon from a particular partial-wave state.
   All entries are in units of MeV.}
   \label{tab_NM}
  \smallskip
\begin{tabular*}{\textwidth}{@{\extracolsep{\fill}}ccccccc}
\hline
\hline
\noalign{\smallskip}
     &  Magic & $\Delta$NNLO(450)$_ {\bf GO}$ & $\Delta$NNLO(394)$_ {\bf GO}$ 
 & $\Delta$NNLO(450)$_{\bf Rf}$ & $\Delta$NNLO(394)$_{\bf Rf}$ &
  \\
\hline  
\hline
\noalign{\smallskip}
  $U(^1P_1)$  &3.71 & 2.93 & 2.29 & 3.71 & 3.74 &  \\
$U(^3P_0)$   & --3.21 & --3.71 & --3.56 & --3.30 & --3.21 &  \\
 $U(^3P_2)$  & --7.81 & --9.49 & --9.15 & --7.62 & --7.79 &  \\
\hline
\noalign{\smallskip}
$U(^1P_1)+U(^3P_0)+U(^3P_2)$ & --7.31 & --10.27 & --10.42 & --7.21 & --7.26 &  \\
\hline
\noalign{\smallskip}
$U(^3S_1)$    & --27.07 & --24.35 & --24.61 & --22.55 &--23.55 &  \\
\hline
\hline
\noalign{\smallskip}
$U$   & --47.63 & --47.11 & --47.56 & --42.25 & --43.11 &  \\
\hline
\noalign{\smallskip}
$T$ & 22.67 & 22.67 & 22.67 & 22.67 & 22.67 & \\
\hline
\noalign{\smallskip}
$E$ & --24.96 & --24.43 & --24.88 & --19.57 & --20.43 &  \\
\hline
\hline
\end{tabular*}
\end{table}

Next, we will explore the impact on nuclear matter from discrepancies in the fit of $NN$ lower partial waves.
To investigate this aspect, we show in Table~\ref{tab_NM} the contributions 
to the nuclear matter energy around saturation density from distinct partial-wave states.
We provide results
for the original GO potentials as well as the refit potentials and the Magic potential.
Of particular interest are the three $P$-waves that we singled out in Table~\ref{tab_ph} and
Fig.~\ref{fig_ph4}.
 It is seen that, for all three $P$-waves, the contributions from the GO potentials are substatially more attractive than
 for the other cases. Recalling that, as demonstrated in 
 Fig.~\ref{fig_ph4} and Table~\ref{tab_ph}, all GO potentials overpredict the empirical
 phase shifts, the increased attraction they generate is not surprising. Thus, while accurately fitted potentials
 obtain about --7.3 MeV from the three $P$-waves, the GO potentials produce 
 about --10.3 MeV, that is 3 MeV more binding energy per particle. Naturally, this is not a viable source for the additional attraction needed in nuclear structure.
 
 The remaining extra attraction by the GO potentials comes from the $^3S_1$ state,
 and is on average $\approx 1.5$ MeV as compared to the corresponding Rf potentials.
 This additional gain in binding energy is, again, linked to unsatisfactory description of phase parameters, in this case the $\epsilon_1$ mixing parameter, cf.\ the $\epsilon_1$ frame in Fig.~\ref{fig_ph1}.
 The explantion of this effect is somewhat involved. 
 Note that the $\epsilon_1$ parameter is proportional to the strength of the nuclear tensor force.
For states in which the tensor force has a dominant role (like the $^3S_1$--$^3D_1$--$\epsilon_1$ system), the $\widehat{T}$-matrix, Eq.~(\ref{eq_LS}), is approximately given by:
 \begin{equation}
 \widehat{T}({\vec p}~',{\vec p})  \approx   \widehat{V}_C ({\vec p}~',{\vec p}) +
\int d^3p''\:
\widehat{V}_T ({\vec p}~',{\vec p}~'')\:
\frac{M_N}
{{ p}^{2}-{p''}^{2}+i\epsilon}\:
\widehat{V}_T ({\vec p}~'',{\vec p}) \, ,
\label{eq_LSapprx}
\end{equation}
where $\widehat{V}_C $ denotes the central force and $\widehat{V}_T$ the tensor force.
The on-shell $\widehat{T}$-matrix is related to the phase-shifts (and observables) 
of $NN$ scattering. Thus, potentials that fit the same phase shifts produce the same 
on-shell $\widehat{T}$-matrix elements. However, that does not imply that the
potentials are the same. As evident from the above equation, the $\widehat{T}$-matrix
is essentially the sum of two terms: the central force term, $\widehat{V}_C $, and the
second order in $\widehat{V}_T$. A potential with a strong $\widehat{V}_T$
will produce a large (attractive) second order term and, hence, go along with a
weaker (attractive) central force; as compared to a weak tensor force potential, where 
the lack of attration by the second order term has to be compensated by a stronger
(attractive) central force. 

Now, when we enter nuclear matter, we encounter particle-particle ladder graphs
represented by
the $\widehat{G}$-matrix:
  \begin{equation}
 \widehat{G}({\vec p}~',{\vec p})  =  \widehat{V} ({\vec p}~',{\vec p}) +
\int d^3p''\:
\widehat{V} ({\vec p}~',{\vec p}~'')\:
\frac{M_N^\star \, Q_P}
{{ p}^{2}-{p''}^{2}}\:
\widehat{G} ({\vec p}~'',{\vec p}) \, ,
\label{eq_G}
\end{equation}
which---similarly to what happened above to the $\widehat{T}$-matrix---for 
states where the tensor force rules,
 can be approximated by:
  \begin{equation}
 \widehat{G}({\vec p}~',{\vec p})  \approx   \widehat{V}_C ({\vec p}~',{\vec p}) +
\int d^3p''\:
\widehat{V}_T ({\vec p}~',{\vec p}~'')\:
\frac{M_N^\star \, Q_P}
{{ p}^{2}-{p''}^{2}}\:
\widehat{V}_T ({\vec p}~'',{\vec p}) \, .
\label{eq_Gapprx}
\end{equation}
The $\widehat{G}$-matrix equation differs from the $\widehat{T}$-matrix equation in two ways: First,
 the Pauli projector, $Q_P$, which prevents scattering into occupied states and, thus, cuts out the low-momentum spectrum. Second, the single-particle spectrum in nuclear matter, which enhances the energy denominator, thereby decreasing the integrand.
 Using a simple parametrization of the single particle energies in nuclear matter, 
 this effect comes down to simply replacing
  the free nucleon mass, $M_N$, by the effective mass $M_N^\star < M_N$.

 Both medium effects reduce the size of the (attractive) integral term and, thus, are repulsive. The larger $V_T$ and the second order $V_T$ term, the larger the repulsive effects. Thus, large tensor force 
 potentials undergo a larger reduction of attraction from these medium effects than weak tensor force potentials.
 This explains the well-known fact that $NN$ potentials with a weaker tensor force yield more attractive results
 when applied in nuclear few- and many-body systems as compared to their strong tensor
 force counterparts.
 
 The GO potentials have a very weak tensor force, which explains their relatively large $^3S_1$ contribution (cf.\ Table~\ref{tab_NM}). In fact, the tensor force is excessively weak, as can be inferred from the underpredicted
 $\epsilon_1$ parameter (cf.\ Fig.~\ref{fig_ph1}).
 To agree with the empirical information within the truncation error, the tensor force
 has to be stronger, like in the case of the Rf potentials, leading to less binding energy in nuclear matter.
 
 At this point of our discussion, a word is in place about what laboratory energies 
 of $NN$ scattering are most relevant for predictions in many-body systems.
 In $P$-waves, about 95\% of the contributions to the $\widehat{G}$-matrix, Eq.~(\ref{eq_G}),
 comes from the Born term. In the sum of the energy contributions from
 zero to the Fermi-momentum $k_F$, the average relative momentum is
 $\bar{p}=\sqrt{0.3} \, k_F$ which, for $k_F=1.35$ fm$^{-1}$, yields $\bar{p}=146$ MeV/c,
 equivalent to $T_{\rm lab}=45$ MeV. Thus, for $P$-waves, the phase shifts around 
 $T_{\rm lab}=50$ MeV are most relevant for nuclear matter predictions at
 saturation density. 
 This fact is particularly evident from the $^3P_0$ phase shifts shown in Fig.~\ref{fig_ph4}.
 In this figure, it is clearly seen that the phase shift predictions around 50 MeV by the
 GO potentials are substantially too large, meaning too attractive.
 On the other hand, the $^3P_0$ phase shifts above 100 MeV by the same potentials (cf.\
 Fig.~\ref{fig_ph1}) are too low, implying too repulsive. But from Table~\ref{tab_NM}
 we know that the GO potentials generate a $^3P_0$ contribution that is too attractive. The conclusion then is
 that phase shifts below 100 MeV are the most relevant ones for many-body predictions
 at normal densities -- the reason why
 we chose $T_{\rm lab} = 50$ MeV for the discussion of $P$-wave phase shift
 in Table~\ref{tab_ph}.
 
 The story is different for states where the tensor force plays a dominant role, like in the
 coupled $^3S_1$--$^3D_1$--$\epsilon_1$ system, where the integral term in
 Eq.~(\ref{eq_Gapprx}) makes a large contribution.
 Note that the integration extends
 from around $k_F$ (due to Pauli blocking) to the cutoff region
 of the potential. Thus, the relative momenta involved 
 are $p \gea k_F$, equivalent to $T_{\rm lab} \gea 150$ MeV for $k_F = 1.35$ fm$^{-1}$,
 which explains why the $\epsilon_1$ mixing parameter needs to be considered for
 energies of 150 MeV or even higher (cf.\ Table~\ref{tab_ph}).
 
 Finally, a comment on the many-body predictions by Magic 
 is in order. As seen in  Table~\ref{tab_NM},
 the $P$-wave contributions from Magic are essentially the same as the ones from 
 the properly fitted Rf potentials, namely --7.31 MeV from the three $P$-waves of special
 interest. 
  What sets the Magic potential apart from all the others is the exceptionally large $^3S_1$
 contribution -- note that the $\epsilon_1$ 
 predictions by Magic are identical to the ones by the N$^3$LO potential of Ref.~\cite{EM03},
 which are right on the data up 300 MeV. 
 The extraordinarily nonlocal nature of Magic due to its
 similarity renormalization group (SRG) evolution is the source of the additional attraction that shows up in nuclear structure.
 This has the consequence that
 the second order $V_T$ term in Eq.~(\ref{eq_Gapprx}) is unusually small
 and, consequently, the central force, $V_C$, unusually large and attractive, giving rise
to the very large, attractive $^3S_1$ contribution by Magic. 
 This degree of nonlocality can,
 presently,
 not be achieved by any original chiral potential, no matter if $\Delta$-full or $\Delta$-less and, therefore,
 these potentials cannot generate $^3S_1$ contributions as large as the Magic one.
  Making up for this by incorrect, extra attractive $P$-waves is not a valid solution.

To summarize, when the three $P$-waves and the $\epsilon_1$ parameter of the GO potentials are corrected to obtain a realistic fit, the favorable predictions for intermediate-mass nuclei are very likely to disappear, as did the extra attraction in nuclear matter.

\section{Conclusions}

We have closely investigated chiral $NN$ potentials at NNLO including $\Delta$-isobar
degrees of freedom and have come to the following conclusions:
\begin{enumerate}
\item
The $\Delta$-full $NN$ potentials at NNLO constructed by the 
G\H{o}teborg-Oak Ridge (GO) group~\cite{Jia20} are up to 40 times outside the 
theoretical error of  chiral EFT at NNLO and are, therefore, inconsistent with the 
EFT that the potentials are intended to be based upon. 
In line with this fact, 
these potentials reproduce the $NN$ data
with a very large $\chi^2$.
This is unacceptable based on contemporary precision standards.
\item
The predictions by the GO $NN$ potentials for the energy per nucleon in nuclear matter
are very attractive, similar to the predictions by 
 the 1.8/2.0(EM) $NN$ potential of Ref.~\cite{Heb11}, also known as `Magic'.
 The extremely attractive nature of both the GO and the Magic potentials
 is the reason for the favorable reproduction of the energies (and radii) of intermediate-mass nuclei,
 which have proven to be a problem in {\it ab initio} nuclear structure physics.
 However, the extra attraction in the GO potentials which brings them to the level of Magic can be traced to incorrect 
  $P$-wave and $\epsilon_1$ mixing parameters. 
\item
When all phase parameters, including the $P$-wave and the $\epsilon_1$-mixing parameters, are fitted within the NNLO truncation error, then the extra attraction disappears and
the nuclear matter predictions become very similar to the ones by $NN$ potentials constructed
within the $\Delta$-less theory.
Thus, we find claims that $\Delta$-full potentials lead to more
attraction in nuclear many-body systems to be incorrect. 
\item
The extraordinarily attractive nature of Magic is due to its high degree of nonlocality which, in turn,
is due to its SRG construction. This degree of nonlocality is not achieved by chiral
$NN$ potentials, no matter if $\Delta$s are included or excluded, because all 
two-pion exchange (2PE) contributions in both version of the theory are local (at least
up to NNLO, see appendix) and nonlocality is generated only by the regulator function,
which adds only moderate nonlocality.
\item
The problem with
a microscopic description of intermediate mass nuclei with realistic chiral nuclear forces remains, unfortunately, unsolved.
\end{enumerate}

 \section*{Acknowledgements}
This work was supported in part by the U.S. Department of Energy
under Grant No.~DE-FG02-03ER41270 (R.M. and Y.N.), 
 by Ministerio de Ciencia e Innovaci\'on under Contract No. PID2019-105439GB-C22/AEI/10.13039/501100011033 and by EU Horizon 2020 research and innovation program, STRONG-2020 project, under grant agreement No 824093 (D.R.E.).

\appendix

\section{The long-range potential}
\label{app_a}

\subsection{Leading order}
\label{sec_lo}

At leading order, only one-pion exchange (1PE) contributes to the long range.
The charge-independent 1PE is given by
\begin{equation}
V_{1\pi}^{\rm(CI)} ({\vec p}~', \vec p) = - 
\frac{g_A^2}{4f_\pi^2}
\: 
\bm{\tau}_1 \cdot \bm{\tau}_2 
\:
\frac{
\vec \sigma_1 \cdot \vec q \,\, \vec \sigma_2 \cdot \vec q}
{q^2 + m_\pi^2} 
\,,
\label{eq_1PEci}
\end{equation}
where ${\vec p}\,'$ and $\vec p$ denote the final and initial nucleon momenta in the 
center-of-mass system, 
respectively. Moreover, $\vec q = {\vec p}\,' - \vec p$ is the momentum transfer, 
 and $\vec \sigma_{1,2}$ and $\bm{\tau}_{1,2}$ are the spin 
and isospin operators of nucleon 1 and 2, respectively. Parameters
$g_A$, $f_\pi$, and $m_\pi$ denote the axial-vector coupling constant,
pion-decay constant, and the pion mass, respectively. See Table~\ref{tab_basic}
for their values.  
Higher order corrections to the 1PE  are taken care of by  mass
and coupling constant renormalizations. Note also that, on 
shell, there are no relativistic corrections. Thus, we apply  1PE in the form
\eq{eq_1PEci} through all orders.

For the $NN$ potentials considered in this paper, the charge-dependence of the 1PE due to pion-mass splitting is taken into account. Thus, in proton-proton ($pp$) and neutron-neutron ($nn$) scattering, we actually use
\begin{equation}
V_{1\pi}^{(pp)} ({\vec p}~', \vec p) =
V_{1\pi}^{(nn)} ({\vec p}~', \vec p) 
= V_{1\pi} (m_{\pi^0})
\,,
\label{eq_1pepp}
\end{equation}
and in neutron-proton ($np$) scattering,
we apply
\begin{equation}
V_{1\pi}^{(np)} ({\vec p}~', \vec p) 
= -V_{1\pi} (m_{\pi^0}) + (-1)^{I+1}\, 2\, V_{1\pi} (m_{\pi^\pm})
\,,
\label{eq_1penp}
\end{equation}
where $I=0,1$ denotes the total isospin of the two-nucleon system and
\begin{equation}
V_{1\pi} (m_\pi) \equiv - \,
\frac{g_A^2}{4f_\pi^2} \,
\frac{
\vec \sigma_1 \cdot \vec q \,\, \vec \sigma_2 \cdot \vec q}
{q^2 + m_\pi^2} 
\,,
\end{equation}
with the exact values for the various pion masses shown in Table~\ref{tab_basic}.  

In this context, we note that, in the 2PE contributions, we neglect the charge-dependence due to
pion-mass splitting and apply $m_\pi = \bar{m}_\pi$ (cf.\ Table~\ref{tab_basic}).

\subsection{Next-to-leading order}
\label{sec_nlo}

We will present the contributions from all subleading pion exchanges in terms of the following template:
\begin{eqnarray} 
V({\vec p}~', \vec p) &  = &
 \:\, V_C \:\, + \bm{\tau}_1 \cdot \bm{\tau}_2 \, W_C 
\nonumber \\ & + &
\left[ \, V_S \:\, + \bm{\tau}_1 \cdot \bm{\tau}_2 \, W_S 
\,\:\, \right] \,
\vec\sigma_1 \cdot \vec \sigma_2
\nonumber \\ &+& 
\left[ \, V_T \:\,     + \bm{\tau}_1 \cdot \bm{\tau}_2 \, W_T 
\,\:\, \right] \,
\vec \sigma_1 \cdot \vec q \,\, \vec \sigma_2 \cdot \vec q  
\, .
\label{eq_nnamp}
\end{eqnarray}

Moreover, we regularize the loop contributions from subleading pion exchanges by spectral-function regularization (SFR) \cite{EGM04}
employing a finite $\tilde{\Lambda} \geq 2m_\pi$.
The purpose of the finite scale $\tilde{\Lambda}$ is to constrain the loop contributions to the  
low-momentum region where chiral effective field theory is applicable.  
Thus, a reasonable choice for $\tilde{\Lambda}$ is to keep it below the masses of the vector mesons
$\rho(770)$ and $\omega(782)$, but above the $f_0(500)$ [also know as $\sigma(500)$]~\cite{PDG}.
This suggests that the region 600-700 MeV is appropriate for $\tilde{\Lambda}$.
For the GO potentials~\cite{Jia20}, $\tilde{\Lambda} =700$ MeV is used, 
while, following Ref.~\cite{EMN17}, $\tilde{\Lambda} =650$ MeV
is applied for the ``Rf'' potentials.

\subsubsection{$\Delta$-less contributions}

The $\Delta$-less $NN$ diagrams that occur at NLO (cf.\ Fig.~\ref{fig_delta_2nf})
contribute in the following way~\cite{KBW97}:
\begin{eqnarray} 
W_C &=&{L(\tilde{\Lambda};q)\over384\pi^2 f_\pi^4} \left[4m_\pi^2(1+4g_A^2-5g_A^4)
+q^2(1+10g_A^2-23g_A^4) - {48g_A^4 m_\pi^4 \over w^2} \right] \,,  
\label{eq_2C}
\\   
V_T &=& -{1\over q^2} V_{S} \; = \; -{3g_A^4 \over 64\pi^2 f_\pi^4} L(\tilde{\Lambda};q)\,,
\label{eq_2T}
\end{eqnarray}  
where the (regularized) logarithmic loop function is given by:
\begin{equation} 
L(\tilde{\Lambda};q) = {w\over 2q} 
\ln {\frac{\tilde{\Lambda}^2(2m_\pi^2+q^2)-2m_\pi^2 q^2+\tilde{\Lambda}\sqrt{
\tilde{\Lambda}^2-4m_\pi^2}\, q\,w}{2m_\pi^2(\tilde{\Lambda}^2+q^2)}}
\label{eq_L}
\end{equation}
with $ w = \sqrt{4m_\pi^2+q^2}$.

\subsubsection{$\Delta$-full contributions}

The $\Delta$-full diagrams at NLO (cf.\ Fig.~\ref{fig_delta_2nf}) are conveniently subdivided
into three groups~\cite{KGW98,KEM07}:
\begin{itemize}
\item
$\Delta$--excitation in the triangle graph:
\beq
\label{del_tri}
W_C = -\frac{h_A^2}{216 \pi^2 f_\pi^4}\,
\left\{ ( 6\Sigma -\omega^2 ) L(\tilde \Lambda;q) +
12 \Delta^2 \Sigma D(\tilde \Lambda;q) \right\}\,;
\eeq
\item
single $\Delta$--excitation in the box graphs:
\beqa
\label{del_sing}
V_{C} &=& -\frac{g_A^2 \, h_A^2}{12 \pi
  f_\pi^4 \Delta }\, (2m_\pi^2 +q^2)^2 \,  A(\tilde \Lambda;q) \,,
  \nn
W_C &=& - \frac{g_A^2 \, h_A^2}{216 \pi^2 f_\pi^4}\, 
\left\{(12 \Delta^2-20 m_\pi^2-11q^2) L(\tilde \Lambda;q) + 6\Sigma^2
  D(\tilde \Lambda;q) \right\} \,,
\nn
V_T &=&  - \frac{1}{q^2} V_S = -\frac{g_A^2 \, h_A^2}{48 \pi^2 f_\pi^4}\,  
\left\{ -2L(\tilde \Lambda;q) + (\omega^2-4 \Delta^2) D(\tilde \Lambda;q) \right\}\,, 
\nn
W_T &=& - \frac{1}{q^2} W_S = -\frac{g_A^2 \, h_A^2}{144 \pi f_\pi^4 \Delta}\,
  \omega^2 \, A(\tilde \Lambda;q)~;
\eeqa
\item
double $\Delta$--excitation in the box graphs:
\beqa
\label{del_doub}
V_{C} &=& -\frac{h_A^4}{27 \pi^2
  f_\pi^4}\, \left\{-4 \Delta^2 L(\tilde \Lambda;q) + \Sigma \left[ H(\tilde\Lambda;q) 
  + (\Sigma +8 \Delta^2)D(\tilde \Lambda;q) \right] \right\}\,, 
  \nn
W_C &=& - \frac{h_A^4}{486 \pi^2 f_\pi^4}\, 
\left\{(12 \Sigma - \omega^2) L(\tilde \Lambda;q) + 3\Sigma \left[ H(\tilde\Lambda;q)
+ (8 \Delta^2 - \Sigma )D(\tilde \Lambda;q) \right] \right\} \,,
\nn
V_T &=&  - \frac{1}{q^2} V_S = - \frac{h_A^4}{216 \pi^2 f_\pi^4}\,  
\left\{ 6 L(\tilde \Lambda;q) + (12 \Delta^2- \omega^2) D(\tilde \Lambda;q) \right\}\,, 
\nn
W_T &=& - \frac{1}{q^2} W_S = - \frac{h_A^4}{1296  \pi^2 f_\pi^4} \,
  \left\{
2L(\tilde \Lambda;q) + (4 \Delta^2 + \omega^2 ) D(\tilde \Lambda;q) \right\}\,;
\eeqa
\end{itemize}
where we are using the following functions:
\beqa
\label{definitions}
\Sigma  &=&  2m_\pi^2 + q^2 -2 \Delta^2\,, 
\nn
A(\tilde{\Lambda};q) &=& {1\over 2q} \arctan{q (\tilde{\Lambda}-2m_\pi) \over q^2
+2\tilde{\Lambda} m_\pi} \,, 
\nn
D(\tilde \Lambda;q) &=& \frac{1}{\Delta} \, \int_{2m_\pi}^{\tilde \Lambda}
\frac{d\mu}{\mu^2+q^2} \, \arctan \frac{\sqrt{\mu^2-4m_\pi^2}}{2\Delta}\,,
 \nn
H(\tilde \Lambda;q) &=& \frac{2 \Sigma }{\omega^2-4 \Delta^2} \left[
  L(\tilde \Lambda;q) - L\left(\tilde \Lambda; 2\sqrt{\Delta^2 -m_\pi^2} \right) 
  \right] \,;
\eeqa
and $\Delta\equiv  M_\Delta - \bar{M}_N$ the $\Delta$-nucleon mass difference (Table~\ref{tab_basic}).
Notice that $\Delta$ is charge-independent to avoid randomly defined charge-dependence.

\subsection{Next-to-next-to-leading order}
\label{sec_nnlo}
\subsubsection{$\Delta$-less contributions}

The $\Delta$-less NNLO contribution (cf.\ Fig.~\ref{fig_delta_2nf}) is given by~\cite{KBW97}:
\begin{eqnarray} 
V_C &=&  {3g_A^2 \over 16\pi f_\pi^4} \left[2m_\pi^2(c_3- 2c_1)+c_3 q^2 \right](2m_\pi^2+q^2) 
A(\tilde{\Lambda};q) \,, \label{eq_3C}
\\
W_T &=&-{1\over q^2}W_{S} =-{g_A^2 \over 32\pi f_\pi^4} c_4 w^2  A(\tilde{\Lambda};q)\,.
\label{eq_3T}
\end{eqnarray}

\subsubsection{$\Delta$-full contributions}

The subleading triangle diagram with $\Delta$ excitation  (cf.\ Fig.~\ref{fig_delta_2nf}) 
makes the following contribution~\cite{KEM07}
(note that we set $(b_3 + b_8) = 0$~\cite{Sie17,Sie20}):
\beqa
\label{del_tri3}
V_{C} &=& -\frac{h_A^2 \, \Delta }{18 \pi^2 f_\pi^4}\, 
\left\{ 6 \Sigma \, \left[ 4 c_1 m_\pi^2 - 2 c_2 \Delta^2 - c_3 ( 2
  \Delta^2 + \Sigma ) \right] \, D(\tilde \Lambda;q) 
  \right.
  \nn
      && \left. + \left[ - 24 \, c_1 m_\pi^2 + c_2 \, 
  (\omega^2 - 6 \Sigma ) + 6 \, c_3 \, ( 2 \Delta^2 + \Sigma ) \right] 
  L(\tilde\Lambda;q) \right\}  \,, 
  \nn
W_T &=& - \frac{1}{q^2} W_S = -\frac{c_4 \, h_A^2 \, \Delta }{72 \pi^2 f_\pi^4}\,
  \left\{ ( \omega^2 - 4 \Delta^2 )  D(\tilde \Lambda;q) 
  - 2 L(\tilde\Lambda;q) \right\} \,. 
\eeqa

\section{The short-range potential}
\label{app_b}

\subsection{Zeroth order}
The zeroth order (leading order, LO) contact potential is given by
\begin{equation}
V_{\rm ct}^{(0)}(\vec{p'},\vec{p}) =
C_S +
C_T \, \vec{\sigma}_1 \cdot \vec{\sigma}_2 \, 
\label{eq_ct0}
\end{equation}
and, in terms of partial waves, 
\be
V_{\rm ct}^{(0)}(^1 S_0)          &=&  \widetilde{C}_{^1 S_0} =
4\pi\, ( C_S - 3 \, C_T )
\nonumber \\
V_{\rm ct}^{(0)}(^3 S_1)          &=&  \widetilde{C}_{^3 S_1} =
4\pi\, ( C_S + C_T ) \,.
\label{eq_ct0_pw}
\ee
To deal with the isospin breaking in the $^1S_0$ state, we treat $\widetilde{C}_{^1 S_0}$
in a charge-dependent way.
Thus, we will distinguish between $\widetilde{C}_{^1 S_0}^{( pp)}$,
$\widetilde{C}_{^1 S_0}^{( np)}$, and $\widetilde{C}_{^1 S_0}^{( nn)}$.

\subsection{Second order}
At second order (NLO), we have
\be
V_{\rm ct}^{(2)}(\vec{p'},\vec{p}) &=&
C_1 \, q^2 +
C_2 \, k^2 
\nonumber 
\\ &+& 
\left(
C_3 \, q^2 +
C_4 \, k^2 
\right) \vec{\sigma}_1 \cdot \vec{\sigma}_2 
\nonumber 
\\
&+& C_5 \left( -i \vec{S} \cdot (\vec{q} \times \vec{k}) \right)
\nonumber 
\\ &+& 
 C_6 \, ( \vec{\sigma}_1 \cdot \vec{q} )\,( \vec{\sigma}_2 \cdot 
\vec{q} )
\nonumber 
\\ &+& 
 C_7 \, ( \vec{\sigma}_1 \cdot \vec{k} )\,( \vec{\sigma}_2 \cdot 
\vec{k} ) \,,
\label{eq_ct2}
\ee
where $\vec k =({\vec p}\,' + \vec p)/2$ denotes the average momentum and $\vec S =(\vec\sigma_1+\vec\sigma_2)/2 $ is the total spin.
Partial-wave decomposition yields
\be
V_{\rm ct}^{(2)}(^1 S_0)          &=&  C_{^1 S_0} ( p^2 + {p'}^2 ) 
\nonumber \\
V_{\rm ct}^{(2)}(^3 S_1)          &=&  C_{^3 S_1} ( p^2 + {p'}^2 ) 
\nonumber \\
V_{\rm ct}^{(2)}(^3 S_1- ^3 D_1)  &=&  C_{^3 S_1- ^3 D_1}  p^2 
\nonumber \\
V_{\rm ct}^{(2)}(^3 D_1- ^3 S_1)  &=&  C_{^3 S_1- ^3 D_1}  {p'}^2 
\nonumber \\
V_{\rm ct}^{(2)}(^1 P_1)          &=&  C_{^1 P_1} \, p p' 
\nonumber \\
V_{\rm ct}^{(2)}(^3 P_0)          &=&  C_{^3 P_0} \, p p'
\nonumber \\
V_{\rm ct}^{(2)}(^3 P_1)          &=&  C_{^3 P_1} \, p p' 
\nonumber \\
V_{\rm ct}^{(2)}(^3 P_2)          &=&  C_{^3 P_2} \, p p' 
\,.
\label{eq_ct2_pw}
\ee
The relationship between the $C_{^{(2S+1)}L_J}$ and the $C_i$ can be found in Ref.~\cite{ME11}.

\section{Definition of nonrelativistic potential}
\label{app_c}

\subsection{Lippmann-Schwinger equation}

The potential $V$ is, in principal, an invariant amplitude (with relativity taken into account perturbatively) and, thus, satisfies a relativistic scattering equation, like, e.\ g., the
Blankenbeclar-Sugar (BbS) equation~\cite{BS66},
which reads explicitly,
\begin{equation}
{T}({\vec p}~',{\vec p})= {V}({\vec p}~',{\vec p})+
\int \frac{d^3p''}{(2\pi)^3} \:
{V}({\vec p}~',{\vec p}~'') \:
\frac{M_N^2}{E_{p''}} \:  
\frac{1}
{{ p}^{2}-{p''}^{2}+i\epsilon} \:
{T}({\vec p}~'',{\vec p}) 
\label{eq_bbs2}
\end{equation}
with $E_{p''}\equiv \sqrt{M_N^2 + {p''}^2}$ and $M_N$ the nucleon mass.
The advantage of using a relativistic scattering equation is that it automatically
includes relativistic kinematical corrections to all orders. Thus, in the scattering equation,
no propagator modifications are necessary when moving up to higher orders.

Defining
\begin{equation}
\widehat{V}({\vec p}~',{\vec p})
\equiv 
\frac{1}{(2\pi)^3}
\sqrt{\frac{M_N}{E_{p'}}}\:  
{V}({\vec p}~',{\vec p})\:
 \sqrt{\frac{M_N}{E_{p}}}
\label{eq_minrel1}
\end{equation}
and
\begin{equation}
\widehat{T}({\vec p}~',{\vec p})
\equiv 
\frac{1}{(2\pi)^3}
\sqrt{\frac{M_N}{E_{p'}}}\:  
{T}({\vec p}~',{\vec p})\:
 \sqrt{\frac{M_N}{E_{p}}}
\,,
\label{eq_minrel2}
\end{equation}
where the factor $1/(2\pi)^3$ is added for convenience,
the BbS equation collapses into the usual, nonrelativistic
Lippmann-Schwinger (LS) equation,
\begin{equation}
 \widehat{T}({\vec p}~',{\vec p})= \widehat{V}({\vec p}~',{\vec p})+
\int d^3p''\:
\widehat{V}({\vec p}~',{\vec p}~'')\:
\frac{M_N}
{{ p}^{2}-{p''}^{2}+i\epsilon}\:
\widehat{T}({\vec p}~'',{\vec p}) \, .
\label{eq_LS}
\end{equation}
Since 
$\widehat V$ 
satisfies Eq.~(\ref{eq_LS}), 
it may be regarded as a nonrelativistic potential. By the same token, 
$\widehat{T}$ 
may be considered as the nonrelativistic 
T-matrix.
All technical aspects associated with the solution of the LS equation 
can be found in Appendix A of Ref.~\cite{Mac01}, including 
specific formulas for the calculation of the $np$ and $pp$ phase shifts (with Coulomb).
Additional details 
concerning the relevant operators and their decompositions 
are given in section~4 of Ref.~\cite{EAH71}. Finally, computational methods
to solve the LS equation are found in Ref.~\cite{Mac93}.

\subsection{Regularization}
\label{sec_reno}

Iteration of $\widehat V$ in the LS equation, Eq.~(\ref{eq_LS}),
requires cutting $\widehat V$ off for high momenta to avoid infinities.
This is consistent with the fact that chiral EFT
is a low-momentum expansion which
is valid only for momenta $Q < \Lambda_\chi \approx 1$ GeV.
Therefore, the potential $\widehat V$
is multiplied
with the (nonlocal) regulator function $f(p',p)$,
\begin{equation}
{\widehat V}(\vec{ p}~',{\vec p})
\longmapsto
{\widehat V}(\vec{ p}~',{\vec p}) \, f(p',p) 
\end{equation}
with
\begin{equation}
f(p',p) = \exp[-(p'/\Lambda)^{2n}-(p/\Lambda)^{2n}] \,.
\label{eq_f}
\end{equation}
In this work, $\Lambda$ is either 450 MeV or 394 MeV.
The exponent $n$ is to be chosen such that the regulator introduces contributions
that are beyond the given order.
In the case of the NNLO potentials of this paper where the given order is three,
this is guaranteed if,  for a contribution of order $\nu$, $n$ is fixed such
that $2n+\nu>3$.
For the GO potentials~\cite{Jia20}, $n=3$ is used 
for $\Lambda = 450$ MeV and $n=4$ for $\Lambda = 394$ MeV.
In the case of the ``Rf'' potentials, we
follow Ref.~\cite{EMN17} and choose $n=2$ for all contributions, except for 
$V_{\rm ct}^{(0)}(^3 S_1) $,
$V_{\rm ct}^{(2)}(^3 P_1) $,   and
$V_{\rm ct}^{(2)}(^3 P_2) $
where $n=3,3$, and 2.5, respectively; and $n=4$ for
1PE.


\begin{thebibliography}{999}
\bibitem{foot1}
The $\chi^2$ for the Hamada-Johnston potential is from Ref.~\cite{SS93}.
The exact bins are 0--125 and 0--183;
and the $NN$ data base is the Nijmegen one~\cite{Sto93}.
\bibitem{foot2}
To calculate the NNLO truncation errors, we have constructed
LO and $\Delta$-full NLO potentials, besides the $\Delta$NNLO ones, 
and applied Eq.~(\ref{eq_error}).
\end{thebibliography}
\end{document}